\documentstyle[psfig]{mn}
\newcommand{\multa}{\mbox{$M_{\rm ext}$}}
\newcommand{\multi}{\mbox{$M_{\rm int}$}}
\newcommand{\vc}{\mbox{$v_{\rm c}$}}
\newcommand{\sigr}{\mbox{$\sigma_{\rm r}$}}
\newcommand{\sigt}{\mbox{$\sigma_{\rm t}$}}
\newcommand{\LCDM}{\mbox{$\Lambda$CDM}}
\newcommand{\TCDM}{\mbox{$\tau$CDM}}
\def\dd{\hbox{\rm d}}
\def\etal{{\rm et al.}}
\def\eg{{\rm e.g.\ }}
\def\etc{{\rm etc.\ }}
\def\ie{{\rm i.e.\ }}
\def\km{{\rm\thinspace km}}
\def\kpc{{\rm\thinspace kpc}}
\def\Mpc{{\rm\thinspace Mpc}}
\def\Msun{\hbox{$\rm\thinspace M_{\odot}$}}
\def\s{{\rm\thinspace s}}
\def\kmps{\hbox{$\km\s^{-1}\,$}}
\def\kmpspMpc{\hbox{$\km\s^{-1}\Mpc^{-1}\,$}}
\def\refindent{\par \noindent \hang}
\def\paper#1#2#3#4#5{\refindent #1, #2, #3, #4, #5}

\def\conf#1#2#3#4#5#6{\refindent #1, #2, in #3, eds, #4. #5, p.~#6}
\def\preprint#1#2{\refindent #1, #2, preprint}
\def\and{, }
\def\AaA{A\&A}

\def\ApJ{ApJ}
\def\ApJL{ApJ}
\def\ApJS{ApJS}
\def\MN{MNRAS}

\begin{document}

\title[The structure of clusters]
{The structure of galaxy clusters in different cosmologies}
\author[P. A. Thomas \etal]
{
Peter A. Thomas$^1$\thanks{Email: p.a.thomas\@sussex.ac.uk},
J\"org M. Colberg$^2$, 
Hugh M. P. Couchman$^3$,
\newauthor{
George P. Efstathiou$^4$\thanks{Current address: Institute of
Astronomy, Madingley Road, Cambridge, CB3 0HA}
Carlos S. Frenk$^5$,
Adrian R. Jenkins$^5$,
}\newauthor{
Alistair H. Nelson$^6$, 
Roger M. Hutchings$^1$,
John A. Peacock$^7$,
}\newauthor{
Frazer R. Pearce$^5$
and
Simon D. M. White$^2$
(the Virgo Consortium)
}\\
{}$^1$Astronomy Centre, University of Sussex, Falmer, Brighton, BN1 9QH\\
{}$^2$Max Planck Instutute f\"ur Astrophysik,
Karl-Schwarzschild-Strasse 1, 85740, Garching bei M\"unchen, Germany\\
{}$^3$Dept.~of Physics \& Astronomy, Univ.~of Western Ontario, London, Ontario,
N6A 3K7, Canada\\
{}$^4$Dept.~of Astrophysics, Nuclear \& Astrophysics Laboratory, Keble
Road, Oxford, OX1 3RH\\
{}$^5$Dept.~of Physics, South Road, Durham, DH1 3LE\\
{}$^6$Dept.~of Physics, College of Cardiff, P.O.Box 913, Cardiff,
CF4 3TH\\
{}$^7$Royal Obsevatory Edinburgh, Blackford Hill, Edinburgh, EH9 3HJ
}

\date{Accepted 1997 ---. Received 1997 ---; in original form 1997 ---}

\maketitle

\begin{abstract}
We investigate the internal structure of clusters of galaxies in
high-resolution N-body simulations of 4 different cosmologies.  There
is a higher proportion of disordered clusters in critical-density than
in low-density universes, although the structure of relaxed clusters
is very similar in each.  Crude measures of substructure, such as the
shift in the position of the centre-of-mass as the density threshold
is varied, can distinguish the two in a sample of just 20 or so
clusters; it is harder to differentiate between clusters in open and
flat models with the same density parameter.  Most clusters are in
a quasi-steady state within the virial radius and are well-described
by the density profile of Navarro, Frenk \& White (1995).
\end{abstract}


\section{INTRODUCTION}

The Virgo Consortium
is an international collaboration whose aim is to
investigate the large-scale structure of the universe via
high-resolution N-body, hydrodynamical simulations.  The ultimate goal
is to relate the observed galaxy distribution to that of the matter
fluctuations in the early universe and in particular to determine the
biases between the two, both in position and velocity, on various scales.

As a first step, we have carried out pure N-body simulations using a
large number of particles (256$^3\approx17\,$million) and four
different Cold Dark Matter (CDM) cosmologies.  Other papers will
report on the clustering evolution of dark matter and dark matter
halos, the topology of large-scale structure, strong and weak
gravitational lensing properties, \etc.

In this paper we concentrate on the structure of galaxy clusters in
our simulations.  Because of our high-resolution, we can trace both
the large-scale topology and the internal structure of the clusters
simultaneously, finding a smaller variation in cluster properties
between different cosmologies than are found in simulations of
isolated clusters.

The main reason for studying the structure of galaxy clusters is to
try to discriminate between different cosmological models.  In
critical-density universes clustering continues to grow to the present
day, whereas in low-density universes it begins to decline after a
redshift $z\sim\Omega_0^{-1}-1$.  This means that clusters in
low-density universes are expected to be dynamically more relaxed and
to have less substructure, steeper density profiles and rounder
iso-density contours.  Early results from simulations of isolated
clusters showed strong variations of all three of the above properties
with the cosmological density parameter, $\Omega_0$ (\eg\ Mohr
\etal~1995, hereafter M95).  We find a much weaker cosmological
dependence.  Although we do indeed find a higher proportion of
disordered clusters in our critical-density simulations, the
properties of relaxed clusters are very similar in all the cosmologies
we examined.

M95 found much steeper density profiles for clusters in low-density
cosmologies than in critical-density ones.  Cen (1994) arrived at a
similar conclusion, but the evidence is rather weak and may instead
reflect a dependence on mass: there are many more massive clusters in
his critical-density simulation than in his open ones.  Subsequent
studies (e.g.\ Jing~\etal~1995, Crone \etal~1997) have shown that,
when measured at radii enclosing the same mean overdensity, the slope
of the density profiles of all clusters are fairly similar and,
although they tend to be slightly steeper in low-density universes
(i.e.\ the clusters are more isolated), this difference is unlikely to
be detectable.

M95 also obtained very few elongated clusters in their N-body,
hydrodynamic simulations of low-density universes but other studies
find most clusters to be flattened in all cosmologies.
Splinter~\etal~(1997) found no change in shape with $\Omega_0$.
Jing~\etal~(1995) found a slight, but significant tendency for rounder
clusters in low-density simulations, and the present work confirms
this result.

The velocity dispersion profiles of clusters are another potential
discriminant of the density parameter.  Jing \& B\"orner (1995) find
that the velocity dispersion declines with radius significantly more
rapidly in low-density than in critical-density universes (however the
effect is hard to measure even with 100 redshifts per cluster).  By
contrast, we find that dynamically relaxed clusters in all our
simulations are well-fit by the model of Navarro, Frenk \& White
(1995), in agreement with Huss, Jain \& Steinmetz (1997).  Moreover,
we do not find a strong cosmological dependence for the anisotropy
parameter as reported in Crone, Evrard \& Richstone (1994).

Substructure (Richstone, Loeb \& Turner 1992; Bartelmann, Ehlers \&
Schneider 1993; Lacey \& Cole 1993) seems to be the best indicator of
$\Omega_0$.  Other measures merely reflect the different abundance of
relaxed clusters whose properties are relatively independent of the
cosmological setting.  Pinkney~\etal~(1996) applied a wide variety of
tests to simulations of merging and isolated clusters and concluded
that there was no single statistic which was guaranteed to detect it:
instead they recommend a battery of a dozen tests, depending upon the
data available and the type of substructure it is desired to detect.

Crone, Evrard \& Richstone (1996) found that a good test for
discriminating between different cosmologies is the shift in the
centre of mass of the matter contained within an iso-density contour
as the contour level is varied and we concur with this (see
Section~\ref{sec:morphology}).  Other powerful indicators are
moments of the squared-density distribution (Dutta 1995) or of the
two-dimensional potential (Buote \& Tsai 1995) or of the surface
density (Wilson, Cole \& Frenk 1996).  One promising new
method for detecting substructure is the `hierarchical clustering
method' which looks for dynamically bound associations of galaxies
(Serna \& Gerbal 1996, Gurzadyan \& Mazure 1997): however, this is
much more complicated to apply than the above tests and so we do not
consider it here.

The above results are somewhat contradictory and confusing for several
reasons: (i) variable mass resolution, (ii) different box-sizes: some
clusters are isolated and some embedded within large-scale structures,
(iii) different power spectra and methods of generating initial
conditions, (iv) different mixtures of gas and dark matter, etc..  The
data presented here provide a large sample of clusters covering about
a decade in mass, drawn from large-scale simulations yet with a
minimum resolution of more than 1000 particles within the virial
radius.  We feel that our cluster sample is much more homogeneous and
representative of the real cluster population than those of previous
studies.

The outline of the paper is as follows.  The simulations and numerical
method are described in Section~\ref{sec:sim}, as is the method of
cluster identification.  The properties of the clusters and various
statistical measures that describe them are presented in
Section~\ref{sec:results} and the prospects for discriminating
between various cosmologies are assessed in Section~\ref{sec:conc}.

\section{METHODOLOGY}
\label{sec:sim}

\subsection{The simulations}

\begin{table}
\caption{Parameters of the 4 simulations, as described in the text.}
\label{tab:param}
\begin{tabular}{lccccccc}
Label& $\Omega_0$& $\lambda_0$& $\Gamma$& $h$& $\sigma_8$& $N$& $m/(h^{-1}\Msun)$\\
\hline
OCDM& 0.3& 0.0& 0.21& 0.7& 0.85& $200^3$& $1.4\times10^{11}$\\
\LCDM& 0.3& 0.7& 0.21& 0.7& 1.30& $256^3$& $6.8\times10^{10}$\\
\TCDM& 1.0& 0.0& 0.21& 0.5& 0.68& $256^3$& $2.3\times10^{11}$\\
SCDM& 1.0& 0.0& 0.50& 0.5& 0.61& $256^3$& $2.3\times10^{11}$
\end{tabular}
\end{table}

We report on the results of 4 simulations in this paper.  Their
properties are chosen to be as similar as possible, while spanning a
range of cosmological parameters as listed in Table~\ref{tab:param}.
The box-size is fixed at $239.5h^{-1}\Mpc$ in each case, where the
Hubble parameter, $h=H_0/100\kmpspMpc$, is chosen so as to give a
sensible age for the Universe (ranging from 11.5\,Gyr for OCDM to
13.8\,Gyr for \LCDM).  The power spectrum is taken to be that
of cold dark matter (CDM), where for three of the runs the shape
parameter, $\Gamma$, is set equal to the physically-motivated
combination $\Omega_0h$.  In the fourth run, \LCDM, we take
$\Gamma=0.21$ so as to give the same spectral shape as in the
low-density models; this set of parameters could occur, for example,
in a decaying-neutrino model (Efstathiou, Bond \& White 1992)

It was our intention to choose the normalisation of the power spectrum,
$\sigma_8$, so as to give approximately the same number of Abell
clusters in each run (see, for example, White, Efstathiou \& Frenk,
1993; Viana \& Liddle, 1996; Eke, Cole \& Frenk, 1996).  In this we
were only partially successful with a range of a factor of four in
abundance; in fact, the normalisation is too large in all runs except
OCDM.  We have since repeated the other three runs with the correct
normalisation but we choose to work with the original simulations in
this paper as this gives a larger cluster sample: the properties of
the individual clusters (as contrasted with their large-scale
distribution) are almost independent of $\sigma_8$.

In run OCDM, the number of particles was only half that in the other
simulations.  Note, however, that the particle mass is intermediate
between that of the other low-density run, \LCDM, and the
critical-density runs.

The gravitational softening in each run was different (because of the
different mass-resolution) but was never larger than $30h^{-1}\kpc$.
Here we will limit our discussion to scales greater than $60h^{-1}\kpc$.

\subsection{The numerical method}

The initial conditions for the simulations were created by perturbing
particles from a glass-like initial state (White 1996; see also
Couchman, Thomas \& Pearce 1995).  The waves were drawn with random
phases and amplitudes from a Gaussian power-spectrum (Bond \&
Efstathiou 1984) using the prescription of Efstathiou \etal~(1985).

The initial state was evolved forward from $z\approx50$ using the
Hydra code of Couchman, Thomas \& Pearce (1995), the parallelisation of
which is discussed in Pearce \& Couchman (1997).  The timestep was set
equal to $\Delta t=\min(0.25dt_a,0.5dt_v)$, where
$dt_a=\min_i(s/a_i)^{1/2}$, $dt_v=\min_i(s/v_i)$, $s$ is the
softening, $a_i$ the acceleration, $v_i$ the speed, and the subscript
$i$ runs over all particles.  The number of timesteps varied from 956
for the \TCDM\ run to just under 1305 in run \LCDM\ that has a smaller
particle mass and softening.  Energy non-conservation in each run, as
measured by the divergence in the Layzer-Irvine integral, was
approximately 1 percent.

\subsection{Cluster identification}

Observationally, clusters of galaxies are rather poorly-defined
objects.  The traditional optical selection criterion depends upon the
surface density of galaxies, but this is prone to projection effects.
In this paper we adopt a more theoretical viewpoint, defining clusters
as sets of particles which exceed a given overdensity threshold.

The usual way to impose a density cut is to group together particles
which lie within a certain distance, or `linking length', $l$ (the
`friends-of-friends' approach: Davis \etal~1985).  If this is set
equal to $b$ times the mean inter-particle separation, ie
$l=b\bar{n}^{-1/3}$, then the associated overdensity is approximately
$2b^{-3}$.  In this paper we adopt a different approach whereby the
density at the location of each dark matter particle is calculated by
smoothing over 32 neighbours using a smoothed particle hydrodynamics
(SPH) algorithm (see \eg\ Hernquist \& Katz 1989).  This has the
advantage of giving a continuous range of densities whose values are
much less subject to Poisson fluctuation than the nearest-neighbour
approach.

To identify clusters we first select those particles whose density
exceeds some threshold, namely 180 times the critical density.  In
order to treat the various cosmologies in the same manner, we use only
a fraction $\Omega_0(256^3/N)$ of the particles, chosen at random, at
this stage (subsequent analysis uses all the particles).
This ensures that each selected particle corresponds to the same
mass, $2.26\times10^{11}h^{-1}\Msun$.  We then create
a minimal spanning tree of the data which can be truncated at a range
of linking-lengths, such as that given above, to divide the particles
into groups.  In practice, any linking length between $(2/\bar{n})^{1/3}$
and a few times this gives the same division, except where two groups
happen to lie very close together.

We define clusters in terms of an overdensity relative to the critical
density because we wish to compare similar objects in each simulation.
Eke, Cole \& Frenk (1996) show that collapsed structures, in the
spherical top-hat approximation, will have lower overdensities
(relative to critical) in universes with $\Omega_0<1$.  As the true
value of the density parameter is unknown, however, we have no way of
allowing for this in the observations.

\begin{table}
\caption{Cluster data: $N_{\rm clus}$---the
number of clusters with mass exceeding $2\times10^{14}h^{-1}\Msun$,
$\sigma$---the mean 1-D velocity dispersion of a
$2\times10^{14}h^{-1}\Msun$ cluster.}
\label{tab:clusters}
\begin{tabular}{lrc}
Label& $N_{\rm clus}$& \hfil$\sigma/\kmps$\\
\hline
OCDM&  90& 720\\
\LCDM& 208& 730\\
\TCDM& 377& 690\\
SCDM&  289& 680
\end{tabular}
\end{table}

In principle we could search for groups down to relatively small
masses.  In this paper, however, we are interested only in rich
clusters and so we choose a mass-cut of $2\times10^{14}h^{-1}\Msun$.
The number of clusters in each simulation using this definition is
listed in Table~\ref{tab:clusters}.  Also shown is the mean 1-D velocity
dispersion of $2\times10^{14}h^{-1}\Msun$ clusters: these are similar
in the different cosmologies which suggests that our selection
criterion is a sensible one.  The corresponding kinetic temperature is
approximately 3\,keV.  The velocity dispersions are slightly higher in
the low-density universes reflecting the fact that clusters in
these cosmologies tend to form at higher redshift and hence be smaller
and have a higher velocity dispersion for a given mass.

\begin{figure}
$$\vbox{
\psfig{figure=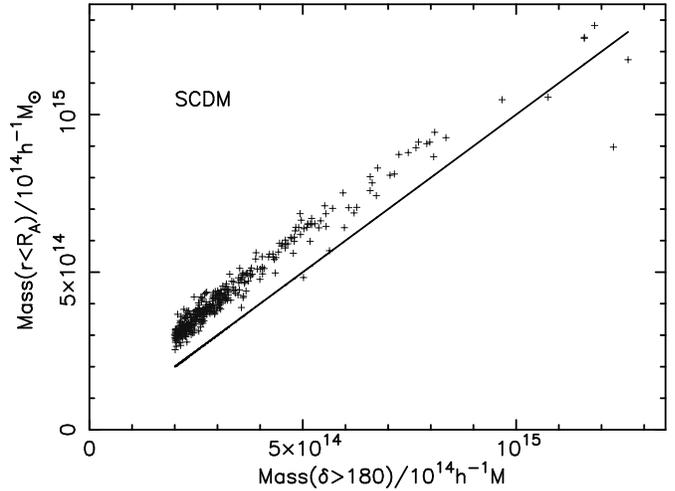,width=8.7truecm,angle=270}
}$$
\caption{The mass of clusters in the SCDM cosmology defined
by two different techniques: abscissa, mass within an overdensity
contour of 180 relative to the critical density; ordinate, mass within
an Abell radius.}
\label{fig:mass}
\end{figure}

An alternative way to define clusters is in terms of the mass
contained within a sphere of fixed radius.  We find a close
correlation between the mass of clusters defined by the overdensity
technique described above and the total mass within an Abell radius
centred on the cluster centroid, which suggests that there is little
difference between the two selection methods.  Figure~\ref{fig:mass}
shows the correspondence for SCDM; the other cosmologies show similar
behaviour.

Throughout this paper we use the term `the mass of the cluster' and
the symbol $m_{180}$ as synonyms for the mass contained within an
overdensity contour of 180 relative to the critical density.

\section{RESULTS}
\label{sec:results}

\subsection{Multiplicity function}
\label{sec:multiplicity}

Theoretically, one of the most basic differences between clusters in
critical and sub-critical density universes is that the latter should
be more relaxed, having formed at a higher redshift.  This should
manifest itself in various ways, one being that these clusters should
be more isolated and have less substructure.
Visually there is little difference between clusters in the different
cosmologies: Figure~\ref{fig:clus} shows some plots of clusters of mass
3--4$\times10^{14}h^{-1}\Msun$ in the OCDM and \TCDM\ runs.
Consequently, we define a statistic to quantify the degree of
isolation of a cluster.

\begin{figure*}
$$\vbox{
\psfig{figure=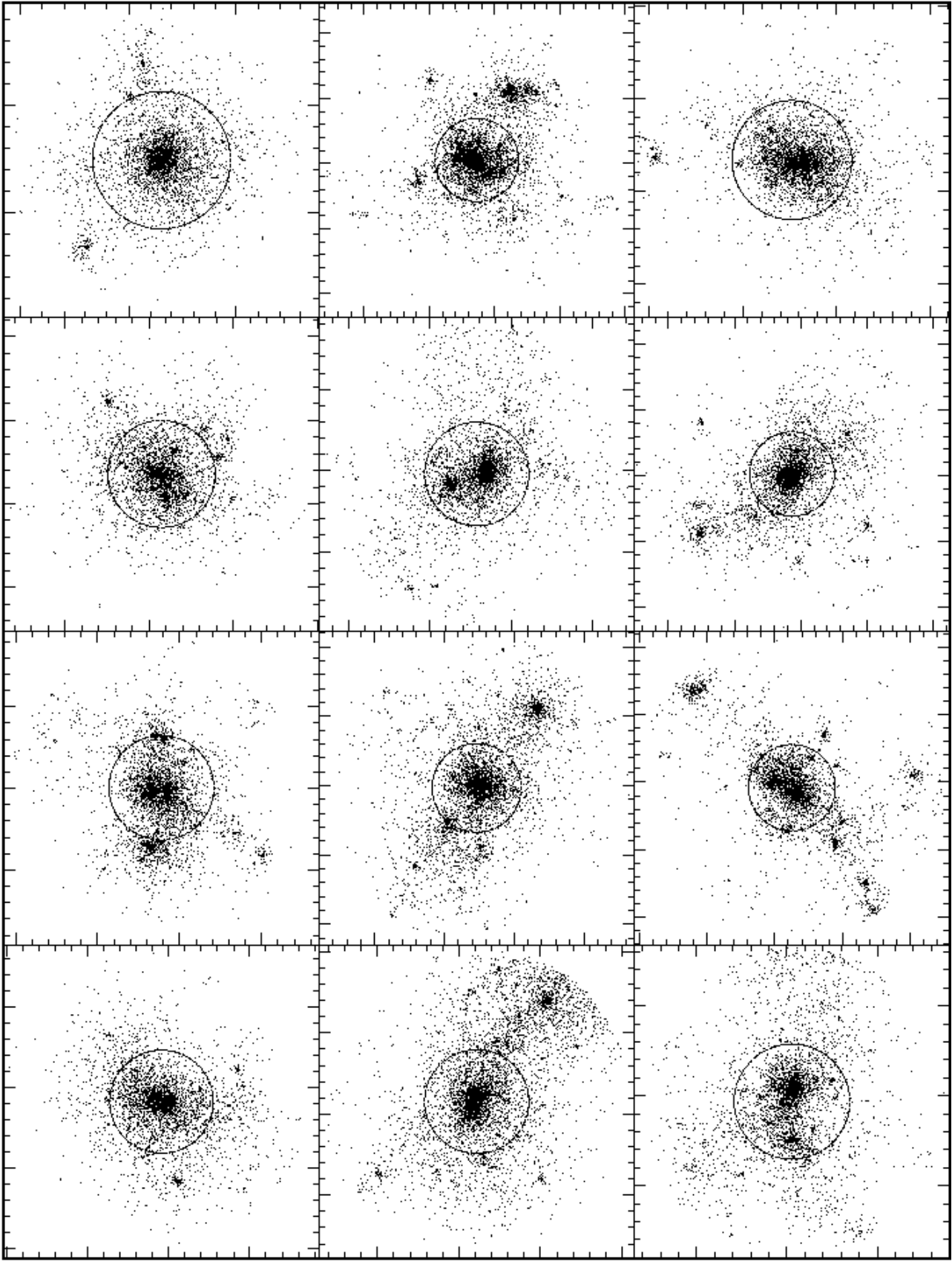,width=17truecm}
}$$
\caption{Example clusters from the OCDM simulation.  The circles
represent an Abell radius.}
\label{fig:clus}
\end{figure*}
\begin{figure*}
$$\vbox{
\psfig{figure=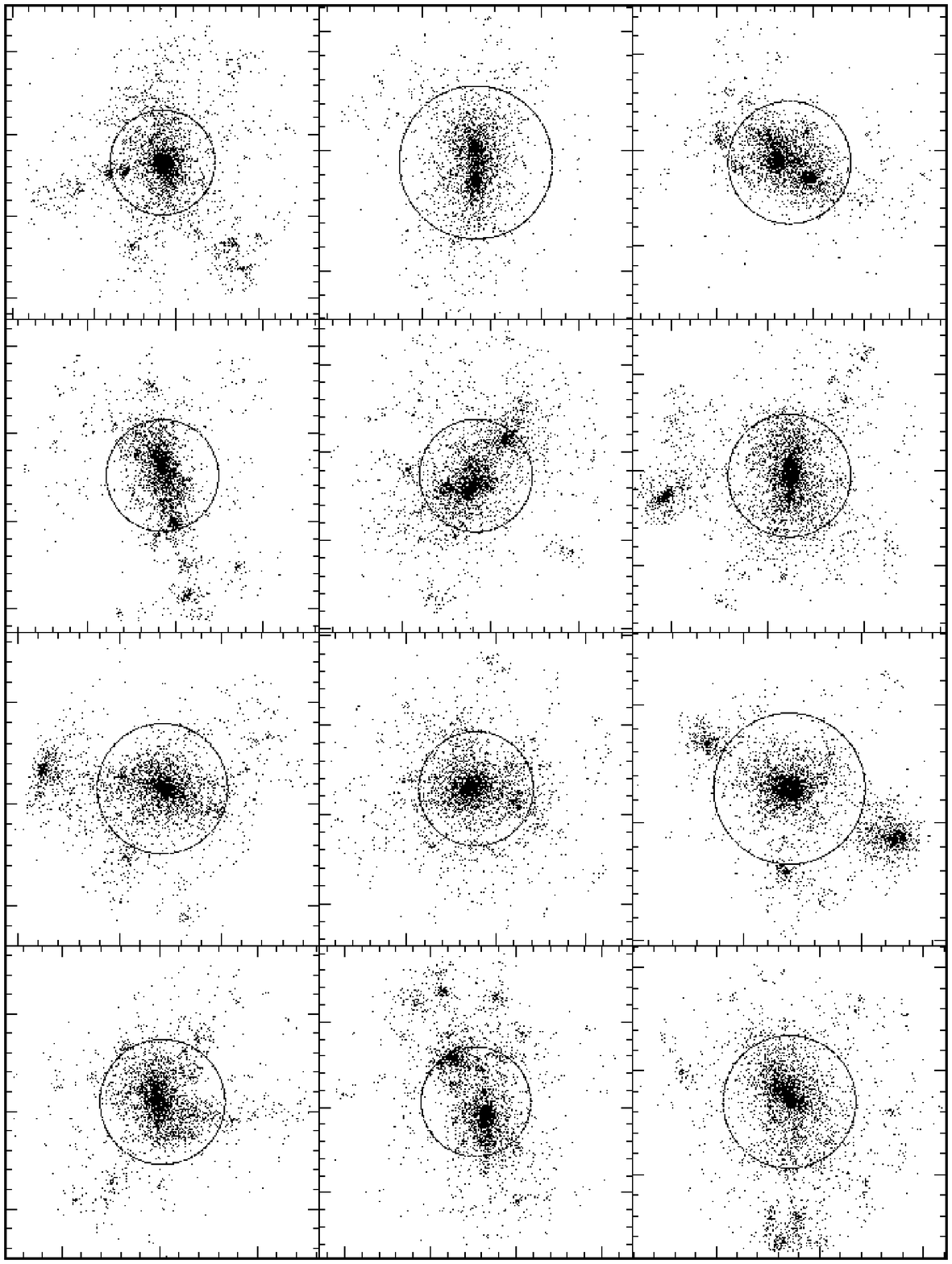,width=17truecm}
}$$
\contcaption{Example clusters from the \TCDM\ simulation.  The circles
represent an Abell radius.}
\end{figure*}

Firstly we catalogue all groups of particles with overdensity greater
than 180 times the critical density and which lie within 2 Abell radii
of the cluster centre (this catalogue is obtained as a by-product of
the method we use to identify clusters).  Then we define a
multiplicity statistic,
\begin{equation}
\multa={M\over m}
\label{eq:multa}
\end{equation}
where $M$ is the total mass of all the groups and $m$ is the mass of
the cluster.

\begin{figure}
$$\vbox{
\psfig{figure=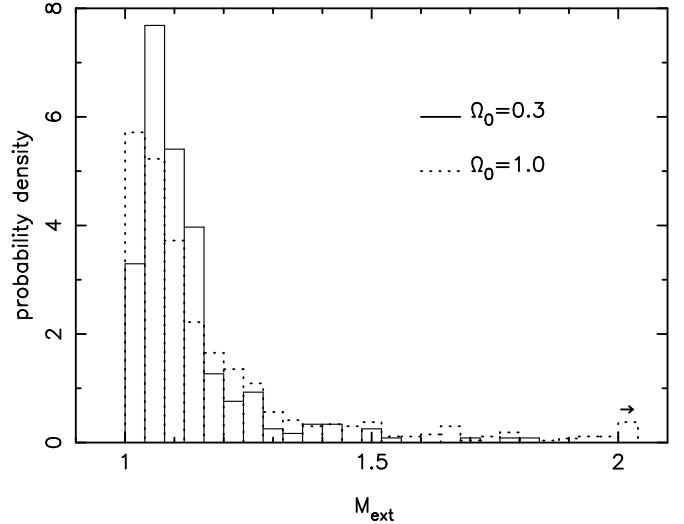,width=8.7truecm,angle=270}
}$$
\caption{Comparison of the probability distributions of the \multa\
statistic for clusters in low-density (solid line) and
critical-density (dotted line) simulations.  All the clusters with
\multa$>2$ are placed in the right-hand bin.}
\label{fig:mla}
\end{figure}

The majority of clusters are relatively isolated, with $\multa<1.2$.
However, there is a small fraction of clusters with larger values of
\multa\ and this fraction increases with decreasing mass.  The two
low-density simulations have identical distributions, as measured by a
Kolmogorov-Smirnov test, as do the two critical-density universes.
\multa\ is plotted against mass for each pair of simulations in
Figure~\ref{fig:mla_13} and a comparison of the two probability 
density distributions is shown in Figure~\ref{fig:mla}.  The
critical-density simulations show a significantly higher fraction of
clusters (approximately one quarter) which have neighbouring
structures with a mass exceeding 20 percent of that of the cluster
itself.  Note, however, that there are few binary clusters (for
which at least one of the pair would have $\multa>2$).  We have
checked that the differences in \multa\ are not due to the different
mass distributions of the two samples but are in fact present at both
low and high masses.  


\begin{figure}
$$\vbox{
\psfig{figure=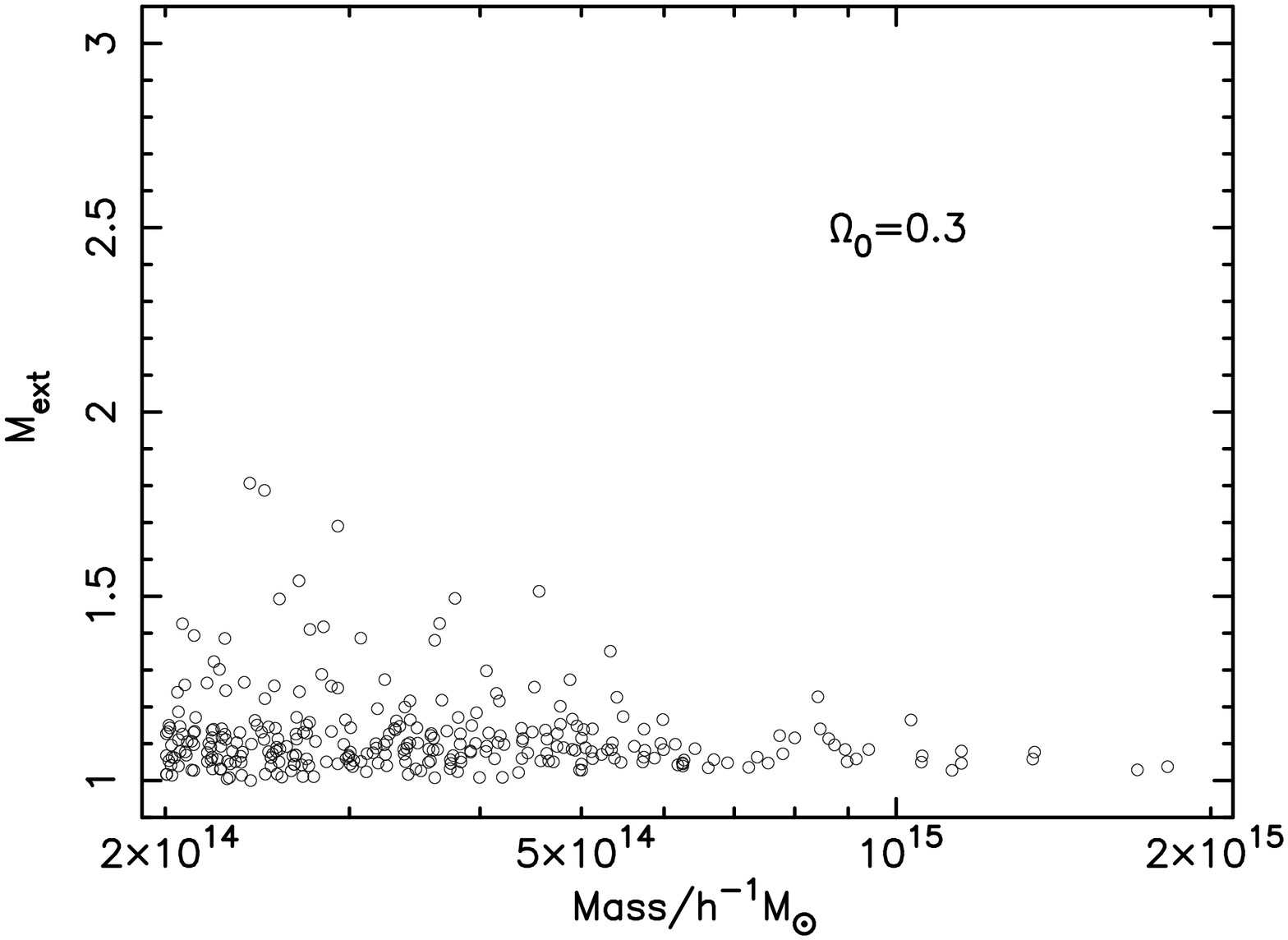,width=8.7truecm,angle=270}
}$$
\caption{Values of the \multa\ statistic for clusters OCDM and \LCDM,
combined.}
\label{fig:mla_13}
\end{figure}
\begin{figure}
$$\vbox{
\psfig{figure=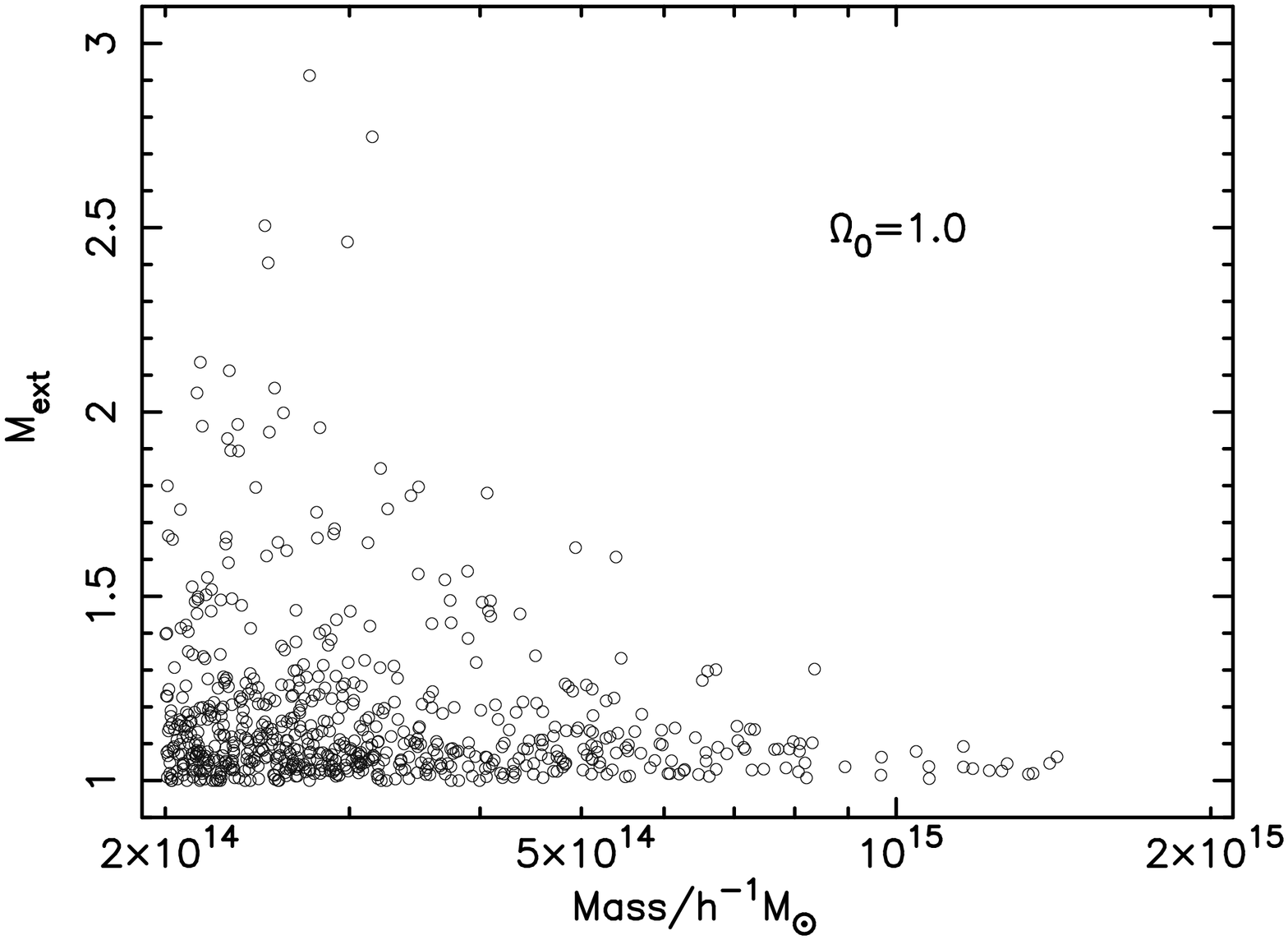,width=8.7truecm,angle=270}
}$$
\contcaption{Values of the \multa\ statistic for clusters \TCDM\ and
SCDM, combined.}
\end{figure}

\subsection{Substructure}
\label{sec:substructure}

As well as being more isolated, clusters in low-density universes are
expected to show less substructure.  We measure this by another
multiplicity statistic.  For each cluster we start with the merger
tree used in its definition and then gradually decrease the linking
length, $l$.  This causes the cluster to break up into subclumps of
higher-and higher overdensity.  At each stage we define
\begin{equation}
\multi={m_1+m_2+m_3\over m_1}
\label{eq:multi}
\end{equation}
where $m_1\geq m_2\geq m_3$ are the masses of the three largest
groups.  The maximum value of \multi\ as $l$ is lowered by a factor
of 4.64 (corresponding to a density increase of a factor of 100) is
used as a measure of the number of subclumps (up to a maximum
of 3) in the catalogue.  We experimented with slightly different
definitions for the multiplicity statistic but all gave similar
results.  

\begin{figure}
$$\vbox{
\psfig{figure=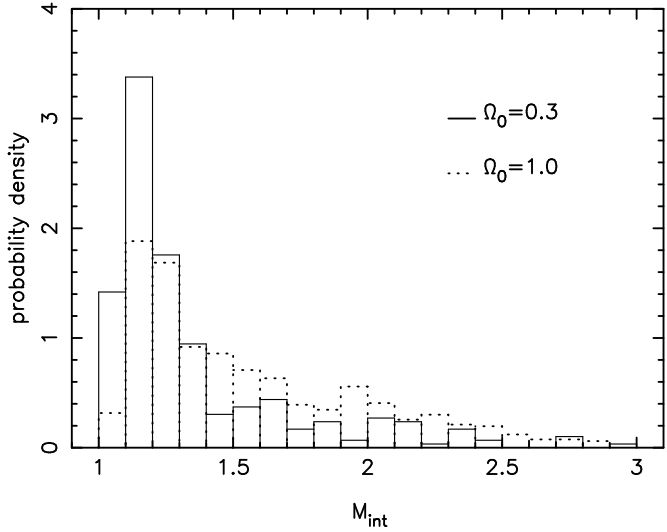,width=8.7truecm,angle=270}
}$$
\caption{Comparison of the probability distributions of the \multi\
statistic for clusters in low-density (solid line) and
critical-density (dotted line) simulations.}
\label{fig:multicomp}
\end{figure}

Once again the two $\Omega_0=0.3$ and the two $\Omega_0=1.0$ universes
give similar distributions, this time with no mass-dependence.  The
probability distributions of the two pairs are compared in
Figure~\ref{fig:multicomp}.  The difference between the two is
significant with the latter showing more substructure.  The mean
values of \multi\ are 1.28 and 1.53, respectively, and clusters in the
critical-density models have twice the frequency of multiple cores
(approximately 40 percent with \multi$>$1.5 and 20 percent with
\multi$>2$).

\begin{figure}
$$\vbox{
\psfig{figure=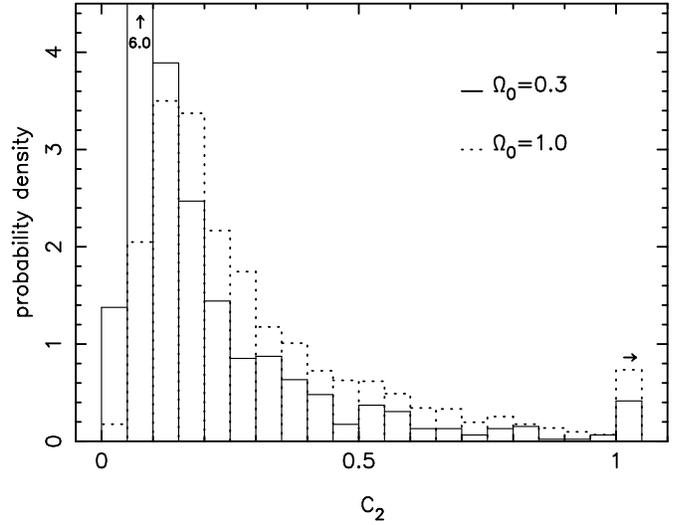,width=8.7truecm,angle=270}
}$$
\caption{Comparison of the probability distributions of the C$_2$
statistic for clusters in low-density (solid line) and
critical-density (dotted line) simulations.  The second bin for
$\Omega_0=0.3$ extends to 6.0; all values of C$_2$ greater than 1.0 are
garnered into the right-hand bin.}
\label{fig:c2}
\end{figure}

In combination, the \multa\ and \multi\ statistics show that there are
very real differences in the degree of substructure between clusters
in low-density and critical-density cosmologies.  The question remains
as to whether these differences would be detectable in practice.
Crone, Evrard \& Richstone (1996) suggested that the shift in the
position of the centre-of-mass of matter contained within an
iso-density contour, as the level of the contour is altered, would be
a good discriminant.  The projected shift, $C_2$ in units of the
cluster radius, $r_{180}$ (defined in Section~\ref{sec:morphology}),
is shown in Figure~\ref{fig:c2} for overdensities in the range
180--18\,000.  With a sample of just 20 clusters, these two
distributions could be distinguished with 5 percent confidence.  A
very similar result is obtained by comparing the position of the
centre-of-mass to that of the density maximum.  Of course, the density
contours of the dark-matter distribution are not directly measurable,
but will be related to those of the X-ray emitting gas in a way which
can be quantified by future simulations.

If one is restricted to observations of galaxies only, then the best
measures of substructure may well be the hierarchical clustering
methods of Serna \& Gerbal (1996) and Gurzadyan \& Mazure (1997).  The
latter find that the existence of 2--3 such subclumps is typical in a
sample of clusters drawn from the ESO Nearby Abell Cluster Survey.
Unfortunately their method is complicated to apply and will be
deferred to a future paper.

\subsection{Morphology}
\label{sec:morphology}

Continuing our theme of looking for evidence of relaxation in clusters
in the low-density cosmologies, we next examine their shapes.  M95
found enormous differences between the axial ratios in
low-density and critical-density universes in their models with the
former being much closer to unity.  We define the semi-axes, 
$a_1\geq a_2\geq a_3$, of each
cluster in terms of the best-fitting ellipsoid, \ie
$a_i=(5\lambda_i)^{0.5}$ where the
$\lambda_i$ are the eigenvalues of the inertia tensor,
\begin{equation}
I_{ij}={1\over m}\sum(r_i-\bar{r}_i)(r_j-\bar{r}_j)
\label{eq:inertia}
\end{equation}
and the sum extends over all $m$ particles in the cluster (the
normalisation is chosen so that for a uniform sphere the semi-axes are
equal to the radius).  The mean radius at which the overdensity equals 180
is $r_{180}=(a_1a_2a_3)^{1/3}$.

\begin{figure}
$$\vbox{
\psfig{figure=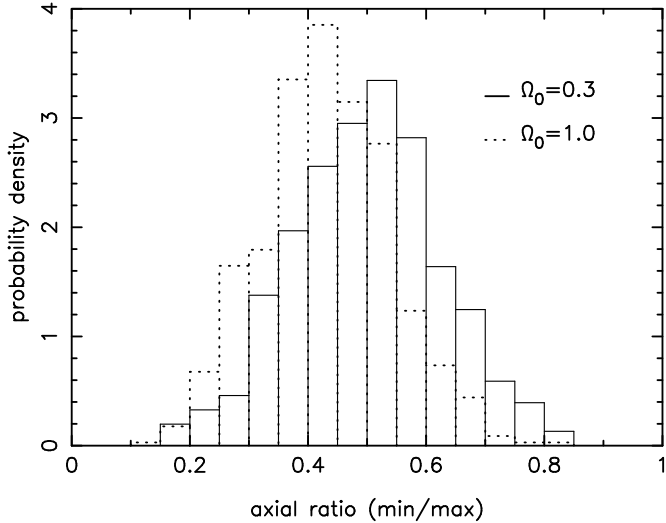,width=8.7truecm,angle=270}
}$$
\caption{Values of the minor/major axis ratio.  $\Omega_0=0.3$ refers
to the combined sample \LCDM\ and OCDM; $\Omega_0=1.0$ to SCDM and \TCDM.}
\label{fig:axial}
\end{figure}

Yet again the distribution of axial ratios in the two low-density
simulations and in the two critical-density simulations are
indistinguishable.  The latter are significantly more elongated (as
measured by a KS test) but only by a small amount.
Figure~\ref{fig:axial} shows a histogram of the minor/major axis ratio
for the two sets of runs.  The mean value of the ratio is 0.50 in the
low-density runs and 0.43 in the high-density runs.  Thus there is a
slight tendency for the clusters in low-density universes to be rounder
than those in critical-density universes, that is consistent with them
being slightly more isolated.  There is also a weaker trend for
high-mass clusters to be rounder than low-mass ones.

\begin{figure}
$$\vbox{
\psfig{figure=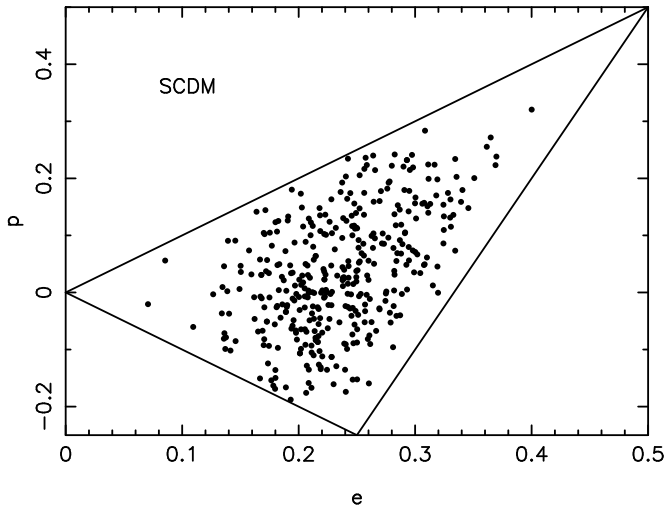,width=8.7truecm,angle=270}
}$$
\caption{Distribution in the ellipticity-prolateness plane of clusters
in the SCDM cosmology.}
\label{fig:ep}
\end{figure}

The three-dimensional shapes of the clusters in all cosmologies show a
wide range of triaxialities, ranging from highly-prolate to
highly-oblate.  Figure~\ref{fig:ep} shows the distribution in
ellipticity-prolateness space for SCDM (all are similar in
appearance). The ellipticity is defined as
\begin{equation}
e={\lambda_3^{-1}-\lambda_1^{-1}\over
2(\lambda_1^{-1}+\lambda_2^{-1}+\lambda_3^{-1})}
\label{eq:ellip}
\end{equation}
and the prolateness as
\begin{equation}
p={\lambda_1^{-1}-2\lambda_2^{-1}+\lambda_3^{-1}\over
2(\lambda_1^{-1}+\lambda_2^{-1}+\lambda_3^{-1})}.
\label{eq:prolate}
\end{equation}

Given that the contours of X-ray emission will be rounder than the
iso-density contours, it is extremely unlikely that observations will
be able to distinguish between low-density and critical-density
universes based simply on the degree of flattening of clusters.  

More sophisticated measures of the morphology of X-ray contours are
based on the moments of the squared-density distribution (Dutta 1995)
or of the two-dimensional potential (Buote \& Tsai 1995).  These are
more closely related to the degree of substructure and so are more
sensitive to the density parameter.  Using the latter method, Buote \&
Xu (1997) claim that SCDM is inconsistent with ROSAT observations of
an, albeit incomplete, sample of bright clusters (Buote \& Tsai,
1996).

We define power ratios in a similar way to Buote \& Tsai (1995):
\begin{equation}
P_m(R_c)={2\over m^2}{(a_m^2+b_m^2)\over a_0^2},
\end{equation}
where
\begin{eqnarray}
a_m &=& \sum_{R<R_c}\rho\,\left(R\over R_c\right)^m\cos{m\phi}\\
b_m &=& \sum_{R<R_c}\rho\,\left(R\over R_c\right)^m\sin{m\phi}.
\end{eqnarray}
Here $(R,\phi)$ are projected polar co-ordinates about the cluster
centroid , $\rho$ is the density and the sum extends over all
particles within 2 Abell radii.

\begin{table}
\caption{KS statistics for distinguishing power ratios within a
0.75$\,h^{-1}$Mpc aperture: $\Delta$ is the maximum difference in
cumulative fraction and $N_{0.05}$ is the number of clusters required
to distinguish the samples with 95 percent confidence using a 1-sided
KS test.  $\Omega_0=1.0$ refers to the combined sample \TCDM\ and SCDM.}
\label{tab:power}
\begin{tabular}{l@{\hspace{0.15cm}}l@{\qquad}c@{\hspace{0.05cm}}c@{\quad}c
@{\hspace{0.05cm}}c@{\quad}c@{\hspace{0.05cm}}c}
&& \multicolumn{2}{c}{$P_2$\hspace*{0.1cm}\hbox{}}&
\multicolumn{2}{c}{$P_3$\hspace*{0.1cm}\hbox{}}& 
\multicolumn{2}{c}{$P_4$\hspace*{0.05cm}\hbox{}}\\
\multicolumn{2}{c}{Samples\hspace*{0.2cm}\ }& $\Delta$& $N_{0.05}$&
$\Delta$& $N_{0.05}$& $\Delta$& $N_{0.05}$\\
\hline
\LCDM& OCDM&        0.122& 125& 0.184&  55& 0.194&  50\\
\LCDM& $\Omega_0=1.0$& 0.206&  45& 0.115& 140& 0.113& 145\\
OCDM& $\Omega_0=1.0$& 0.309&  20& 0.238&  35& 0.268&  25
\end{tabular}
\end{table}

As might be expected the values of the power ratios are lowest for
OCDM (least substructure) and greatest for the two critical density
models, \TCDM\ and SCDM.  There is a slight tendency for high mass
clusters to show more substructure, but this effect is smaller than
the cosmological differences.  The difference in the cumulative
distributions of $P_m(0.75\,h^{-1}\Mpc)$ and the corresponding number
of clusters required to distinguish the different cosmologies using a
KS test are shown in Table~\ref{tab:power}; the critical-density
models have been combined as they are indistinguishable.  In contrast
to Buote \& Xu (1997), we find that $P_2$ is best able to distinguish
low-density from critical-density models, whereas $P_3$ and $P_4$ do
better at discriminating between \LCDM\ and OCDM.  Of course the
numbers in Table~\ref{tab:power} assume a direct correspondence
between the theoretical distribution of dark matter and the observed
X-ray flux.  This association needs to be checked and a practical
application of the method is likely to require more clusters than is
indicated here.

We have also calculated the first moment, $P_1$, within an aperture
centred on the density maximum.  This gives similar results to $P_2$
because they both measure the dipole moment of the cluster.

\begin{figure}
$$\vbox{
\psfig{figure=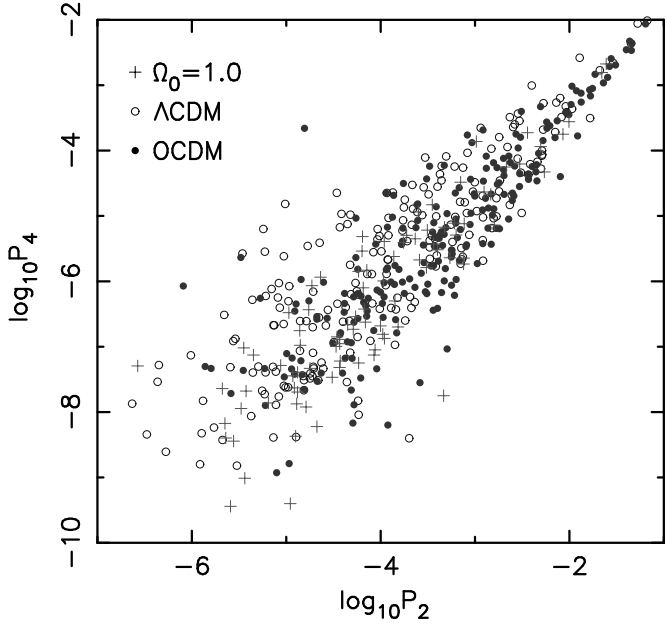,width=8.7truecm,angle=270}
}$$
\caption{$P_2$ versus $P_4$ in $0.75\,h^{-1}$Mpc apertures for the
different cosmologies.  Only a subset of all clusters is shown, for clarity.}
\label{fig:pow}
\end{figure}

The different moments, or power ratios, are highly correlated.  The
strongest relation, shown in Figure~\ref{fig:pow}, is between $P_2$
and $P_4$.  Interestingly the $\Omega_0=1$ cosmologies and OCDM show a
tighter correlation than \LCDM, that seems to extend to higher
values of $P_4$ for a given $P_2$.  This may be another way to
discriminate between models with and without a cosmological constant.

When we repeated the above tests for power ratios defined within an
aperture of variable size, $R_c=r_{180}$, proportional to the virial
radius of the cluster, then the differences between clusters in the
\LCDM\ and OCDM cosmologies were reduced, but the distinction between
these and the critical-density universes was maintained.

\subsection{Density profiles}
\label{sec:dprof}

Our final measure of the degree of isolation of a cluster is the slope
of the density profile.  We use the mean slope, $s$, between the
radius, $r_{180}$, within which the mean density is 180 times the
critical density (defined in Section~\ref{sec:morphology}), and 0.1
times this radius.  Assuming a power-law density distribution
$\rho\propto r^{-s}$ in this range, then $s$ is defined implicitly as
\begin{equation}
{m(r)-m(0.1r)\over m(0.32r)-m(0.1r)}={10^{3-s}-1\over10^{(3-s)/2}-1}
\label{eq:slope}
\end{equation}
where $m(r)$ is the mass within radius $r$.  

\begin{figure}
$$\vbox{
\psfig{figure=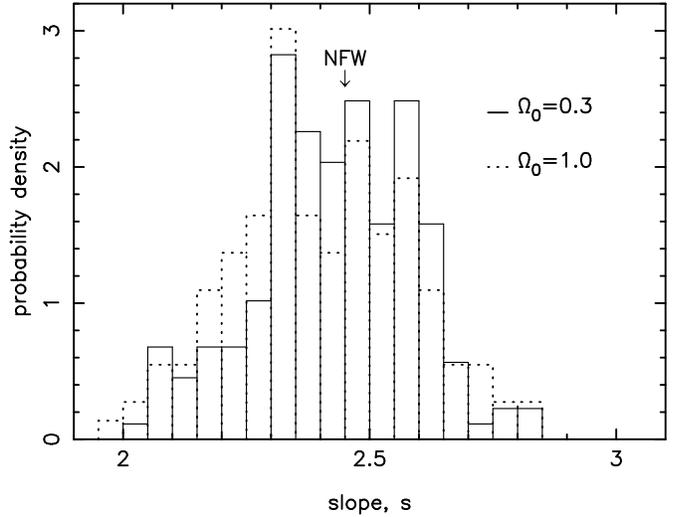,width=8.7truecm,angle=270}
}$$
\caption{Values of the slope of the density profile, as defined in
Equation~\ref{eq:slope}, for clusters with \multi$<$1.2.  The arrow
shows the mean slope of the Navarro, Frenk \& White (1995) profile in
this range.}
\label{fig:slope}
\end{figure}

$s$ is well-defined only for clusters with a fairly smooth density
distribution and so we restrict the analysis to those with
$\multi<1.2$.  Histograms of the distributions of $s$ for low-density
and critical-density runs are shown in Figure~\ref{fig:slope}.  The
two are marginally distinguishable: assuming perfect data one would
need a sample of 90 clusters to distinguish the two at 5 percent
confidence.  The mean slopes are 2.45 and 2.40, respectively,
consistent with our earlier conclusion that clusters in low-density
universes are slightly more isolated (remember too that there are more
clusters with substructure in the critical-density universes which
have been omitted from the analysis).

These differences are less than have been found in previous work.
M95 found $s$ to be much larger in open, low-density
cosmologies than in flat ones, whereas Jing \etal~(1995) found a
weaker effect, but still larger than ours.  We all agree that, for a
given density parameter, clusters in open cosmologies are more
isolated (i.e.\ have steeper density profiles) than those in flat
cosmologies.

There is a weak trend of decreasing $s$ with increasing mass in
low-density cosmologies and increasing $s$ with increasing mass in
critical-density cosmologies.  Thus low-mass clusters show greater
differences in their slope than do high-mass ones.  As our simulations
are of larger volumes, and hence contain more high-mass clusters than
previous studies, this may partly explain the smaller variation in $s$
that we find.

The mean values of $s$ are similar to that of the density profile of
Navarro, Frenk \& White (1995), shown by the arrow in Figure~\ref{fig:slope}.
This will be discussed further in the next section.

Using a fixed physical radius, rather than one defined in terms of
overdensity is much less useful.  Because in more massive clusters one
is measuring the slope at a smaller fraction of the cluster radius,
and because the density profile steepens with increasing radius, there
is a decrease in $s$ with increasing mass.  The measured slope then
becomes strongly dependent upon the mass of the cluster sample.

It is inconceivable that the differences in slope, being much less
than the spread in slopes even for the subset of smooth clusters,
could be used to discriminate between different cosmological models.

\subsection{Velocity dispersion profiles}
\label{sec:vprof}

Figure~\ref{fig:vprof} shows averaged velocity profiles for the
clusters in cosmologies \LCDM\ and SCDM (the others are similar).  In
each case we have included only relatively smooth clusters with
\multi$<1.2$ and have centred the profiles on the density maximum. The
radii are scaled to the radius, $r_{180}$ (defined in
Section~\ref{sec:morphology}), at which the density drops to 180 times
the critical density, and all clusters are given the same weighting in
the average.  For an isothermal profile $r_{180}$ is a fraction
$1/\surd{3}\approx0.58$ of the virial radius (within which the mean
overdensity is 180).

\begin{figure}
$$\vbox{
\psfig{figure=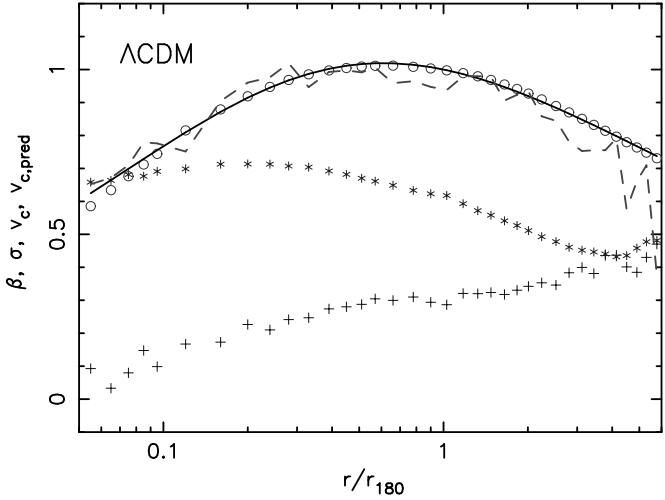,width=8.7truecm,angle=270}
}$$
\caption{Velocity profiles for the combined sample of 100 \LCDM\
clusters with \multi$<1.2$: circles---circular speed, solid line---fit
to circular speed, dashed line---predicted circular speed from
steady-state Jeans' Equation, stars---velocity dispersion,
crosses---anisotropy parameter.}
\label{fig:vprof}
\end{figure}
\begin{figure}
$$\vbox{
\psfig{figure=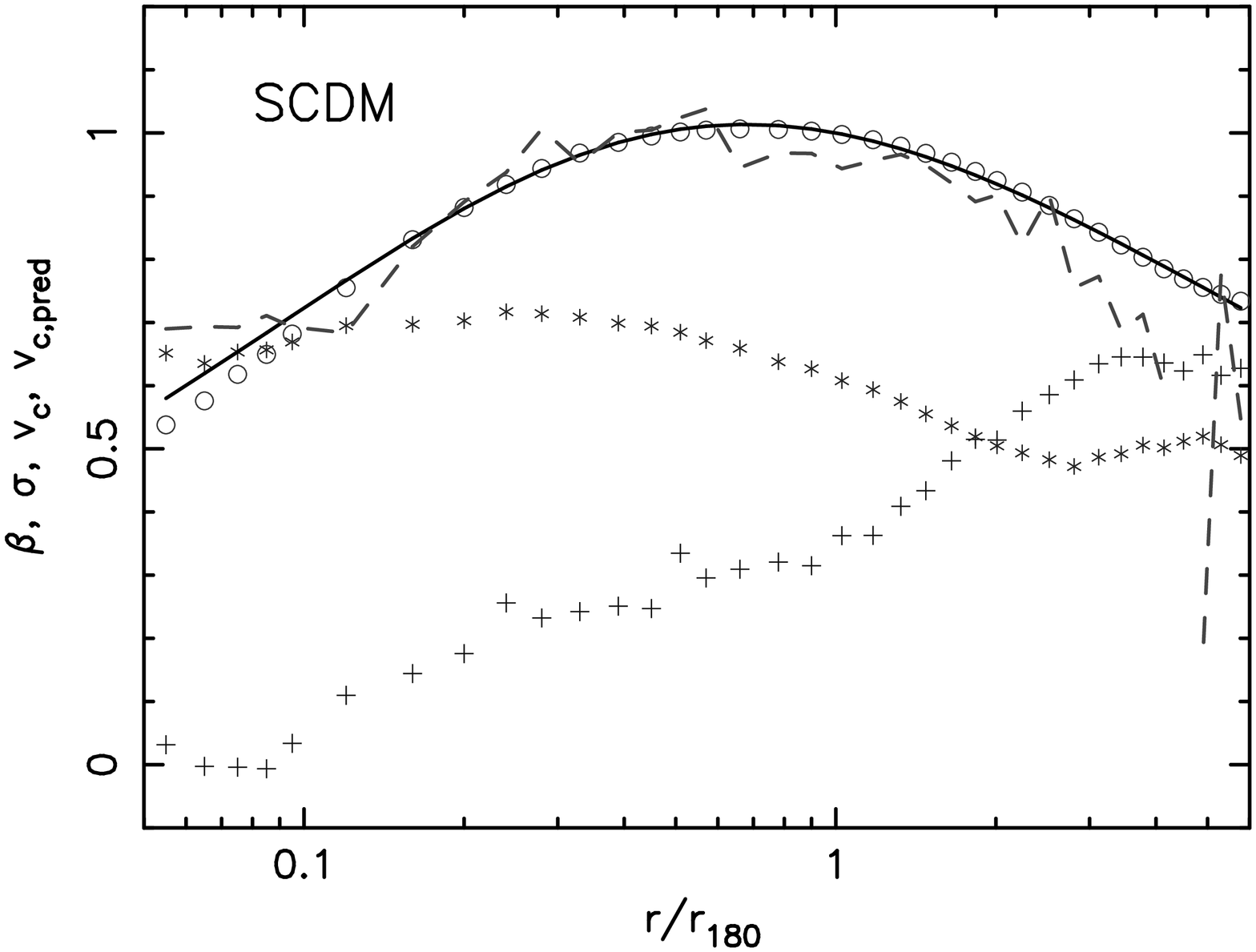,width=8.7truecm,angle=270}
}$$
\contcaption{Velocity profiles for the combined sample of 97 SCDM
clusters with \multi$<1.2$.  Legend as for \LCDM\ plot.}
\end{figure}

The circles in the Figure show the normalised circular velocity
profiles, $\vc=(m\,r_{180}/m_{180}r)^{0.5}$ where $m$ is the mass
contained within radius $r$ and $m_{180}$ is the total mass within
$r_{180}$.  These have been fit by a function of the form
\begin{equation}
\vc^2={x_{180}(1+x_{180})^{s-1}\big[(1+x)^{s-1}-(s-1)x-1\big]\over
x\,(1+x)^{s-1}\big[(1+x_{180})^{s-1}-(s-1)x_{180}-1\big]},
\label{eq:vc2}
\end{equation}
($s\neq2$) corresponding to a density profile
\begin{equation}
\rho\propto{1\over x\,(1+x)^s},
\end{equation}
where $x=r/a$, $x_{180}=r_{180}/a$ and the characteristic radius $a$
and slope $s$ are parameters of the fit.  For $s=2$,
Equation~\ref{eq:vc2} must be modified slightly to
\begin{equation}
\vc^2(s=2)={x_{180}(1+x_{180})\,\big[(1+x)\ln(1+x)-x\big]\over
x\,(1+x)\,\big[(1+x_{180})\ln(1+x_{180})-x_{180}\big]}.
\end{equation}
$s=2$ corresponds to the profile suggested by Navarro, Frenk
\& White (1995, hereafter NFW; 1996) to be a good fit to the galaxy
clusters in their simulations, and $s=3$ to the model proposed by
Hernquist (1990) for the halos of elliptical galaxies.  The
low-density models were fit over a range $0.1<r/r_{180}<4$ and the
critical-density ones over a range $0.1<r/r_{180}<3$.  The lower
radius in each case corresponds to about 60$\,h^{-1}$kpc, comfortably
larger than the softening, and the outer radius was determined by the
break in the velocity dispersion profile.  The fits are shown in
Table~\ref{tab:vc} and by the solid line in the Figure.

\begin{table}
\caption{Fits to the circular velocity profiles}
\label{tab:vc}
\begin{tabular}{lrccc}
Label& $N_{\rm clus}$& $\langle r_{180}\rangle/h^{-1}$kpc& $s$& $a/r_{180}$\\
\hline
OCDM&  54& 0.62& $2.02\pm0.01$& $0.29\pm0.04$\\
\LCDM& 100& 0.56& $1.82\pm0.01$& $0.22\pm0.01$\\
\TCDM&  65& 0.62& $2.01\pm0.03$& $0.37\pm0.01$\\
SCDM&   97& 0.57& $1.99\pm0.02$& $0.31\pm0.03$
\end{tabular}
\end{table}

In each case the extrapolated fit overestimates the circular speed
within $r=0.1\,r_{180}$ which implies that the inner density profile is
shallower than $1/r$; this may, however, be due to the finite
softening which ranges up to 30\,$h^{-1}$kpc for the critical-density
runs.  The clusters in all but one cosmology follow the NFW profile
closely over the range in which they are fitted: the \LCDM\ clusters
would prefer a slightly shallower slope at $r\approx r_{180}$ but
steepen at large radii to agree with the others.  

The `concentration parameter', $c\approx\surd{3}\,r_{180}/a$, is around
6, in rough agreement with the values found by NFW for halos of this
mass.  The low-density cosmologies have higher values of $c$ reflecting
the higher redshift of cluster formation.

The normalised velocity dispersion, $\sigma=\langle
v^2r_{180}/3Gm_{180}\rangle^{1/2}$, is shown by the stars in the
Figure.  In each case it rises gently out to
$r/r_{180}\approx$0.2--0.3, then declines at larger radii.  The
variation in $\sigma^2$ within the virial radius is about a factor of
three; similar temperature variations are seen in hydrodynamic
simulations (e.g.\ Thomas \& Couchman 1992).  Jing \& B\"orner (1995)
found a significant trend for steeper velocity dispersion profiles in
open cosmologies when measured at constant {\it physical} radii.  We
suspect that this trend may disappear if the profiles are normalised
to $r_{180}$, as in Figure~\ref{fig:vprof}.  However, the effect is
anyway quite weak for $\Omega_0=0.3$ and so we will not pursue it
here.

The variation of velocity dispersion with radius is not equal in each
component.  The anisotropy parameter, $\beta=1-\sigt^2/\sigr^2$, where
$\sigr$ and $\sigt$ are the radial and (1-D) tangential components of the
velocity dispersion respectively, is shown by the crosses in
Figure~\ref{fig:vprof}.  The orbits are approximately isotropic in the
core of the cluster but become increasingly radial as one moves out in
radius.  Contrary to the report in Crone, Evrard \& Richstone (1994),
we find no marked difference in the behaviour of the anisotropy
parameter between different cosmologies.

It is usual to treat the central regions of clusters as in a
quasi-steady state.  We test this assumption by checking to see if the
clusters satisfy the spherically-symmetric, steady-state Jeans
Equation:
\begin{equation}
{1\over\rho}{\dd(\rho\sigr^2)\over\dd r}+{2\beta\sigr^2\over r}
=-{\vc^2\over r}.
\end{equation}
The value of \vc\ predicted by this equation is shown by the dashed
line in Figure~\ref{fig:vprof}.   
It is clear that the steady-state assumption is a reasonable one out
to the virial radius.

\subsection{Angular momentum}
\label{sec:angmom}

We measure the angular momentum of the clusters in terms of the
dimensionless spin parameter,
\begin{equation}
\lambda={JT^{1/2}\over Gm^{5/2}},
\label{eq:spin}
\end{equation}
where $J$ is the magnitude of the angular momentum, $T$ is the
kinetic energy and $m$ is the mass of the cluster.  The original
definition (Peebles 1971) uses the total energy of the system rather
than the kinetic energy, but this is much harder to determine and the
two are equivalent for a virialised system.

\begin{figure}
$$\vbox{
\psfig{figure=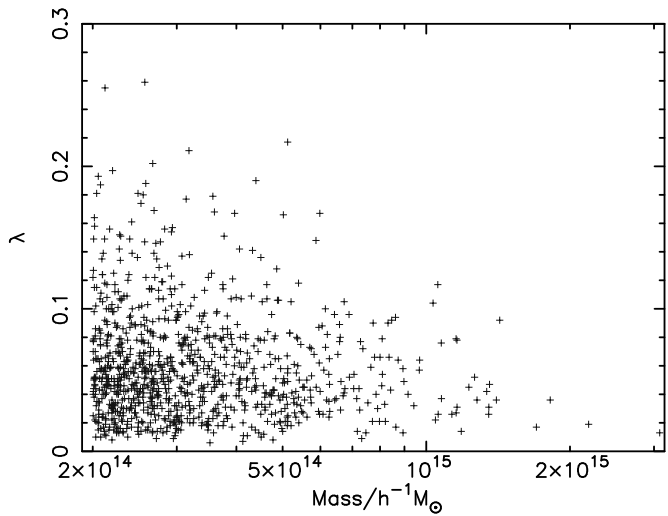,width=8.7truecm,angle=270}
}$$
\caption{Values of the spin parameter for the combined sample of clusters.}
\label{fig:spin}
\end{figure}

The distributions of $\lambda$ are indistinguishable for each of our
cosmologies and the combined sample is shown in Figure~\ref{fig:spin}.
The mean value of $\lambda$ is 0.060 (the 10, 50 and 90 percentiles
are 0.020, 0.051 and 0.108) but there is a significant trend of
increasing mean and variance in $\lambda$ as one moves to smaller
masses.

\section{CONCLUSIONS}
\label{sec:conc}

We have examined the structural differences between clusters in
simulations of four cosmological models.  Our basic conclusion is that
relaxed clusters have very similar properties in all cosmologies, but
that there is a greater proportion of dynamically young clusters in
critical-density universes than in low-density ones.  It is hard to
discriminate between low-density universes with and without a
cosmological constant.

The best way to constrain the density parameter is to look for
substructure in the cluster population.  Thus, in
Section~\ref{sec:substructure} it was shown that clusters in
critical-density simulations are much more likely to have multiple
cores, and that a very simple test, the centre-of-mass shift as the
density contour is varied, can distinguish the two in a population of
just 20 clusters.  The dipole moment of the squared-density
(i.e.\ the $P_2$ power ratio, Section~\ref{sec:morphology}) requires
a similar sample size.  

Secondary measures of substructure, such as the shape of the
iso-density contours, perform less well.  The ellipticities we see (in
agreement with other recent work) are substantially smaller than M95
originally suggested.  This may, in part, be due to our larger boxes.
Most of the clusters which we find appear to be located at the
intersections of large-scale filaments: in this sense none of them are
truly isolated and many are still forming today even in low-density
universes.

The density and velocity-dispersion profiles of clusters have both
been suggested in the past as cosmological probes.  However, we have
shown in Sections~\ref{sec:dprof} and \ref{sec:vprof} that the
profiles of relaxed clusters are very similar in all models we
considered.  They are in a quasi-steady state within the virial radius
(that encloses a mean overdensity of 180), and are well-fit by the
density profile of NFW.  If we look at the complete sample of
clusters, then the mean density and velocity dispersion profiles
decline less rapidly in critical-density models that are less isolated
and have less well-defined centres.  However, we feel that this is a
secondary effect and that it is better to look for substructure
directly.

Some of the properties that we have investigated (such as the degree
of isolation, the major/minor-axis ratio and the angular momentum)
have a slight mass-dependence.  Others, (such as the density and
velocity-dispersion profiles) show more regularity when measured in
units of the virial radius, rather than a fixed physical size.  This
suggests that some of the variation in cluster properties which have
been reported in earlier work may be due to a variation in the
multiplicity function in different cosmologies.  The redshift
evolution of the number density of clusters is also expected to show a
strong cosmological dependence.  This two topics will be investigated in
subsequent papers.

\section*{ACKNOWLEDGMENTS}

The simulations described in this paper were performed on Cray T3Ds at
the RZG of the Max-Planck-Society in Garching and at the EPSC in
Edinburgh as part of the Virgo Project, using an N-body only version
of the Hydra N-body, hydrodynamics code (Couchman, Thomas \& Pearce
1995; Pearce \& Couchman 1997; collaboration supported by NATO
CRG\,970081).  This paper was prepared using the facilities of the
STARLINK minor node at Sussex.  During its production, PAT was partly
supported by a Nuffield Foundation Science Research Fellowship.  CSF
acknowledges a PPARC Senior Research Fellowship.

\section*{REFERENCES}
\paper{Bartelmann M., Ehlers J.\and Schneider P.}{1993}{\AaA}{280}{351}
\paper{Bond R.\and Efstathiou G. P.}{1984}{\ApJ}{285}{L45}
\paper{Buote D. A.\and Tsai J. C.}{1995}{\ApJ}{452}{522}
\paper{Buote D. A.\and Tsai J. C.}{1996}{\ApJ}{458}{27}
\paper{Buote D. A.\and Xu G.}{1997}{\MN}{284}{439}
\paper{Cen R.}{1994}{\ApJ}{437}{12}
\paper{Couchman H. M. P., Thomas P. A.\and Pearce, F. R.}{1995}
{\ApJ}{452}{797}
\paper{Crone M. M., Evrard A. E.\and Richstone D. O.}{1994}
{\ApJ}{434}{402}
\paper{Crone M. M., Evrard A. E.\and Richstone D. O.}{1996}
{\ApJ}{467}{489}
\paper{Crone M. M., Govertano F., Stadel J.\and Quinn T.}{1997}
{\ApJL}{477}{L5}
\paper{Davis M., Efstathiou G. P., Frenk C. S.\and White S. D. M.}
{1985}{\ApJ}{292}{371}
\paper{Dutta S. N.}{1995}{\MN}{276}{1109}
\paper{Efstathiou G. P., Bond J. R.\and White S. D. M.}{1992}{\MN}{258}{1P}
\paper{Efstathiou G. P., Davis M., Frenk C. S.\and White S. D. M.}
{1985}{\ApJS}{57}{241}
\paper{Eke V. R., Cole S.\and Frenk C. S.}{1996}{\MN}{282}{263}
\preprint{Gurzadyan V. G.\and Mazure A.}{1997}\ \AaA, submitted.
\paper{Hernquist L.}{1990}{\ApJ}{356}{359}
\paper{Hernquist L.\and Katz N.}{1989}{\ApJS}{70}{419}
\preprint{Huss A., Jain B.\and Steinmetz M.}{1997}\ Astro-ph/9703014
\paper{Jing Y. P.\and B\"orner G.}{1995}{\MN}{278}{321}
\paper{Jing Y. P., Mo H. J., B\"orner G.\and Fang L. Z.}{1995}
{\MN}{276}{417}
\paper{Lacey C.\and Cole S. M.}{1993}{\MN}{262}{627}
\paper{Mohr J. J., Evrard A. E., Fabricant D. G.\and Geller M. J.}
{1995}{\ApJ}{447}{8}\ {\bf M95}
\paper{Navarro J. F., Frenk C. S.\and White S. D. M.}{1995}
{\MN}{275}{720}\ {\bf NFW}
\paper{Navarro J. F., Frenk C. S.\and White S. D. M.}{1996}{\ApJ}{462}{563}
\preprint{Pearce F. R.\and Couchman H. M. P.}{1997}\ Astro-ph/9703183
\paper{Peebles P. E. J.}{1971}{\AaA}{11}{377}
\paper{Pinkney J., Roettiger K., Burns J. O.\and Bird C. M.}{1996}
{\ApJS}{104}{1}
\paper{Richstone D., Loeb A.\and Turner E. L.}{1992}{\ApJ}{393}{477}
\paper{Serna A.\and Gerbal D.}{1996}{\AaA}{309}{65}
\paper{Splinter R. J., Melott A. L., Linn M. L., Buck C.\and Tinker
J.}{1997}{\ApJ}{479}{79}
\paper{Thomas P. A.\and Couchman H. M. P.}{1992}{\MN}{257}{11}
\paper{Viana P. T. P.\and Liddle A. R.}{1996}{\MN}{281}{323}
\conf{White S. D. M.}{1996}{Cosmology and Large-scale Structure}
{Schaeffer R., Silk J., Spiro M.\and Zinn-Justin J.}{Elsevier}{349}
\paper{White S. D. M, Efstathiou G. P.\and Frenk C. S.}{1993}{\MN}{262}{1023}
\paper{Wilson G., Cole S. M.\and Frenk C. S.}{1996}{\MN}{280}{199}

\end{document}